\documentclass[aps,showpacs,twocolumn, prc, amssymb]{revtex4}
\usepackage{graphics}
\usepackage{graphicx}
\usepackage{amsmath, amscd, amssymb}
\newcommand{\be}{\begin{equation}}
\newcommand{\ee}{\end{equation}}
\newcommand{\bc}{\begin{center}}
\newcommand{\ec}{\end{center}}

\newcommand{\bqr}{\begin{eqnarray}}
\newcommand{\eqr}{\end{eqnarray}}

\begin{document}

\title{{\vglue -0.4 true cm}Observation of the lowest energy $\gamma$-ray in any superdeformed nucleus : $^{196}$Bi}

\author{A.~Pr\'evost}
\altaffiliation{Corresponding author; Present Address: CSNSM Orsay, IN2P3/CNRS, Universit\'e Paris-Sud, F-91405 Orsay Campus, France}
\email[e-mail address : ]{prevost@csnsm.in2p3.fr}
\author{B.~Ross\'e}
\author{M.~Meyer} 
\author{N.~Redon}
\author{C.~Schmitt}
\author{O.~St\'ezowski}
\author{P.~Lautesse}
\affiliation{IPN Lyon, IN2P3/CNRS, Universit\'e Lyon-1, F-69622 Villeurbanne Cedex, France}
\author{A.~Astier}
\author{I.~Deloncle} 
\author{M.G.~Porquet}
\affiliation{CSNSM Orsay, IN2P3/CNRS, Universit\'e Paris-Sud, F-91405 Orsay Campus, France}

\author{ T.~Duguet}
\affiliation{Physics Division, Argonne National Laboratory, Argonne, IL 60439, USA}

\author{ H.~H\"ubel}
\author{E.~Mergel}
\author{D.~Ro{\ss}bach}
\author{N.~Nenoff}
\author{G.~Sch\"onwa{\ss}er}
\affiliation{Helmholtz-Institut f\"ur Strahlen- und Kernphysik, Universit\"at Bonn, Nussallee 14-16, D-53115 Bonn, Germany}

\author{D.~Curien}
\author{G.~Duch\^ene}
\author{B.J.P.~Gall}
\author{J.~Robin}
\author{J.P.~Vivien}
\affiliation{IReS Strasbourg, IN2P3/CNRS, Universit\'e Louis Pasteur, F-67037 Strasbourg, France}

\begin{abstract}New results on the superdeformed $^{196}$Bi nucleus are reported. We have observed with
the EUROBALL IV $\gamma$-ray spectrometer array a superdeformed transition of 124 keV which is the lowest
observed energy $\gamma$-ray in any superdeformed nucleus. We have developped microscopic cranked Hartree-Fock-Bogoliubov
calculations using the SLy4 effective force and a realistic surface pairing which strongly support the
$K^\pi=2^-$($\pi$[651]1/2$\otimes \nu$[752]5/2) assignment of this superdeformed band.

 \end{abstract}


\maketitle

While impressive results exist on the superdeformation phenomenon on Au,
Hg, Tl and Pb isotopes, very scarce data are available beyond Z=82 in the A$\sim$190 mass region.
Only four yrast superdeformed (SD) bands are known in the $^{195,196,197}$Bi \cite{clark1,clark2} and $^{198}$Po\cite{McNabb}
isotopes.\\

In order to deepen our knowledge of the valence SD orbitals in Z=83, high spin states of the $^{196}$Bi nucleus
 have been reinvestigated with the EUROBALL IV $\gamma$-ray spectrometer \cite{simpson_deux}.
 In this brief report, we present the first result obtained in $^{196}$Bi, a 124 keV $\gamma$-ray
 transition which is the lowest $\gamma$-ray ever observed in any SD nuclei with the exception of the fission
 isomers. This line is interpreted as the $9^-\rightarrow 7^-$ transition of the $K^\pi=2^-$
 yrast SD band in the framework of self-consistent lattice microscopic calculations.

\indent The SD states of $^{196}$Bi were populated in the $^{19}$F($^{184}$W,7n) reaction at a beam energy of
114~MeV. The beam was delivered by the VIVITRON accelerator at the IReS laboratory in Strasbourg. The target consisted of
three $^{184}$W foils of thickness 200, 125 and 125 $\mu$g/cm$^{2}$ mounted on thin carbon supports. The
$\gamma$-rays were detected with the EUROBALL IV spectrometer ~\cite{simpson_deux} which comprised an inner-ball
of 210 BGO crystals and 71 Compton-suppressed Ge detectors. Our automatic procedure based on the
 fuzzy set theory \cite{paper_fuzzy} has been applied for the energy calibration of the corresponding 239 Ge
 crystals. A condition of four unsuppressed Ge detectors firing in coincidence combined with an inner-ball multiplicity
 more than 7 was required to record events on DLT tapes. After presorting (prompt-time window,
add-back of the Clover and Cluster composite detectors, Compton rejection), the data set finally consists of
2$\times 10^9$ three- and higher-fold events.

 \begin{figure}[b]

\includegraphics[width=8cm,angle=0]{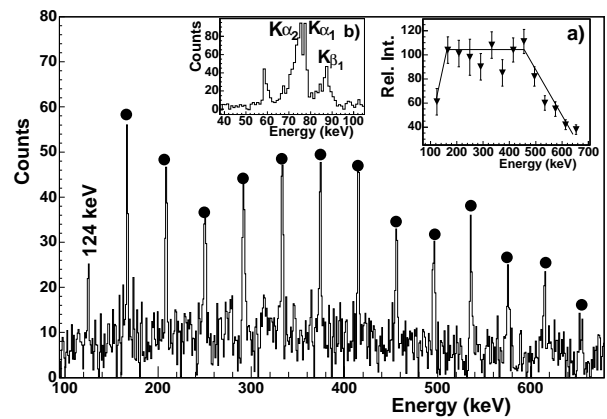}

\label{spectra}

\caption{Four-gated spectrum obtained with our EUROBALL IV experiment showing the new 124 keV transition. The dots indicate the
positions of the other SD transitions. The relative intensity pattern of the yrast SD band in $^{196}$Bi and the region of the spectrum
around the X-rays are plotted in the inserts a) and b).}
\end{figure}

In fig. 1, we present a four-gated spectrum obtained in this experiment. The energies of the SD transitions obtained 
are presented in table 1. We confirm the 13 transitions observed by Clark et al. 
\cite{clark1, clark2} with GAMMASPHERE in double-gated spectra.
The most striking new result is the presence of a new SD transition at the bottom of the band with 
energy of 124.0 keV. The intensity pattern for this band, corrected for internal conversion assuming E2
character for all the transitions, is presented in the insert a) of fig. 1. It is worth noting that the measurement of 
the 124 keV intensity allows us to exhibit an extremely rapid de-excitation in only one transition  after the plateau 
 (with an intensity decrease of roughly 40 \%).

\begin{table}
\label{table_energies}
\caption{Energies, relative intensities and spins proposed for the $^{196}$Bi yrast SD band transitions. A 100 \% intensity is obtained for the plateau.}
\begin{ruledtabular}
\begin{tabular}{ccc}
$E_{\gamma}$ (keV) & I$_{tot}$(\%) & Spin ($\hbar$)\\
\hline
\bf{124.0(3)} & \bf{62(11)} & \bf{($9)\rightarrow (7$)} \\
165.7(4) & 104(10) & ($11)\rightarrow( 9$)  \\
207.6(4) & 100(10) & ($13)\rightarrow (11$) \\
249.5(4) & 98(15) & ($15)\rightarrow (13$) \\
290.9(4) & 90(10) & ($17)\rightarrow (15$) \\
332.6(4) & 108(11) & ($19)\rightarrow (17$) \\
373.5(4) & 85(11) & ($21)\rightarrow (19$)\\
413.3(4) & 104(10) & ($23)\rightarrow (21$) \\
454.6(5) & 111(10) & ($25)\rightarrow (23$) \\
494.8(5) & 82(8) & ($27)\rightarrow (25$) \\
535.1(5) & 60(6) & ($29)\rightarrow (27$) \\
574.9(5) & 55(6) & ($31)\rightarrow (29$) \\
614.0(6) & 42(4) & ($33)\rightarrow (31$) \\
654.2(8) & 38(4) & ($35)\rightarrow (33$) \\
\end{tabular}
\end{ruledtabular}
\end{table}

Despite a large statistic, no normally deformed (ND) transition is observed in coincidence with the SD band.
However the insert b) of fig. 1 presents the low energy part of the spectrum showing that the SD transitions are in coincidence with the X-rays of bismuth
($K_{\alpha_1}$=77.107 keV,$K_{\alpha_2}$=74.815 keV, $K_{\beta_1}$=87.349 keV). We finally adopt the assignment to the
$^{196}$Bi given by Clark et al. \cite{clark1,clark2}.

\begin{figure}[b]
 \includegraphics[width=7.5 cm,angle=0]{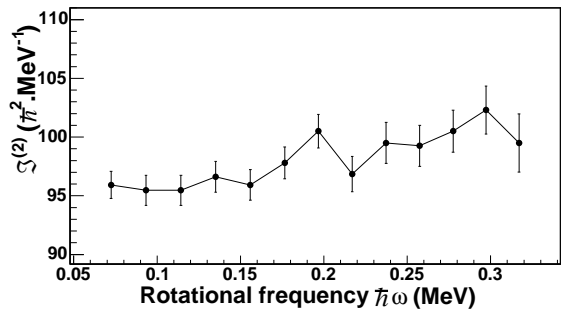}
\label{inertie1}
\caption{Dynamical moment of inertia $\Im^{(2)}$ of the yrast SD band in $^{196}$Bi.}
\end{figure}

\indent The new 124 keV transition extends the flat
behaviour of the dynamical moment of inertia $\Im^{(2)}$ of the yrast SD band in $^{196}$Bi down to very low
frequencies as illustrated in fig. 2.

\indent We have applied the approach of Deloncle et al. \cite{deloncle} to estimate the maximum spin
transferred in the recoiling nucleus. In A$\sim$190 mass region, the variation of the maximum spin observed
in a SD band versus the Z$^2$/A fissility parameter is roughly linear with a change of around 9 per unit
of Z$^2$/A. By an extrapolation to $^{196}$Bi we obtain a maximum spin reached of 35$\pm 2 \hbar$. Considering the
high sensitivity of our measurements the 654 keV transition is certainly the uppest transition 
(no higher-energy transition does exist) and corresponds to a $35 \hbar \rightarrow 33\hbar$ transition. 
The lowest spin is then
expected to be 7$\pm 2 \hbar$. The resulting spin values included in table 1 are in agreement with the spin assignment
obtained by the Harris \cite{harris} and Wu \cite{wu} approaches.
A carefull search of other SD bands in our data has succeeded in retrieving several SD bands in Tl, Pb isotopes and one
in $^{197}$Bi populated according to the PACE predictions \cite{gav}. However no excited SD band in $^{196,197}$Bi has been observed.

\begin{figure*}
\includegraphics[width=16.5cm,angle=0]{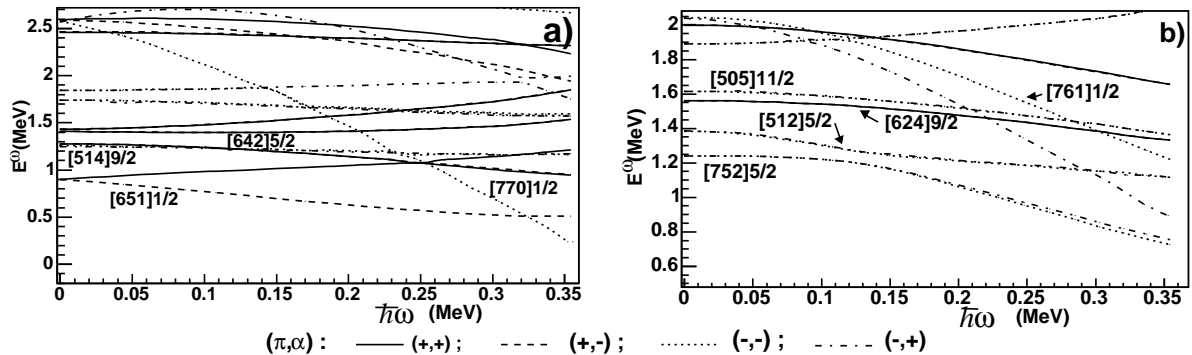}
\caption{Quasi-particle routhians $E^\omega$ of $^{198}$Po as function of the rotational frequency for protons (a) and neutrons (b) obtained with cranked Hartree-Fock-Bogoliubov calculations.}
\end{figure*}

In addition, we have performed to interpret the data static and cranked Hartree-Fock-Bogoliubov (HFB) calculations applied to the adjacent even-even $^{198}$Po
(Z=84 and N=114) nucleus.  These lattice calculations have been performed using the SLy4 parametrization of the Skyrme
nucleon-nucleon effective interaction for the particle-hole channel \cite{Chabanat_publi}. The pairing correlations (the particle-particle channel)
have been taken into account via a zero range density-dependent interaction localized at the surface of the 
nucleus with the same parameters used for heavy nuclei of A$\sim$150 and A$\sim$250 mass regions \cite{Bertsch, terasaki,Rigollet, duguet}.
To restore partially the particle number symmetry the Lipkin-Nogami prescription has been used. Our calculations have been
performed in a tridimensional lattice and the size of the box was 40$\times$40$\times$40 ~fm$^3$ with a mesh
of 0.8~fm.\\
Our static calculations localize the SD minimum at a mass quadrupole moment $Q_{20}$ of around 
5200 fm$^2$ (charge quadrupole moment $Q_{2c}$ of 2270 e.fm$^2$, deformation parameter
$\beta\sim$0.6) similar to the result obtained using the D1S Gogny Force \cite{Gogny}. The quasi-particle
routhians shown in fig. 3 have then been calculated for this minimum.  It appears that,
at low frequency, the lowest available configuration is one quasi-particle (qp) proton in the $\pi$[651]1/2 orbital coupled
to one qp neutron in the $\nu$[752]5/2 orbital.

\indent At low frequencies, in the framework of the strong coupling scheme, the configuration $\pi$[651]1/2$\otimes \nu$[752]5/2 gives rise to two
doublets of bands with K$^\pi=2^-$ and $3^-$, the first one being favoured by the Gallagher-Moskowski rule
\cite{rule_GM}. However, in our experiment we have observed no excited band which would correspond to the signature partner. The Coriolis
coupling between the K$^\pi=2^-$ and K$^\pi=3^-$ bands, the mixing of other configurations, and also the residual
proton-neutron interaction could be responsible of the lowering of the favoured signature partner of the 
K$^\pi=2^-$ band. This band coming from $\pi$[651]1/2$(\alpha=-1/2)\otimes \nu$[752]5/2$(\alpha=-1/2)$ has a total signature $\alpha_{tot}=-1$ 
and an odd spin sequence (3$^-$,5$^-$,7$^-$,...).\\
This interpretation based on our self-consistent mean-field calculations is in accordance with the results
obtained with phenomenologic cranked Wood-Saxon calculations \cite{clark1,clark2} in which the [752]5/2
orbital exhibits an immediate splitting of the two signature partners, the
$\alpha=-1/2$ partner being favoured as the rotational frequency increases. In our calculations
the splitting of the [752]5/2 orbital is less pronounced (the two signature partners are nearly
degenerate up to high frequencies) and we cannot totally reject an interpretation based on the 
$\pi$[651]1/2$(\alpha=-1/2)\otimes \nu$[752]5/2$(\alpha=1/2)$ configuration for the yrast SD band 
leading to the even spin sequence I=(2$^-$,4$^-$,6$^-$,...). However we have finally adopted that the 
favoured configuration remains the $\pi$[651]1/2$(\alpha=-1/2)\otimes \nu$[752]5/2$(\alpha=-1/2)$ with an odd spin
sequence for the SD band. 

\indent This imposes the last SD transition we have observed in $^{196}$Bi, namely the 124 keV line, 
to be the $9^-\rightarrow 7^-$.
Two supplementary transitions (highly converted) should exist to reach the 3$^-$ SD state of the favoured signature of the K$^\pi$=2$^-$ band. 
A similar situation occurs in $^{194}$Pb \cite{waely} where the lowest proposed SD
transition was $6^+\rightarrow 4^+$ compared to the SD band head of 0$^+$. \\

\indent In conclusion, in order to study the superdeformation phenomenon above $Z=82$, we have populated the $^{196}$Bi
nucleus in an experiment with the EUROBALL IV array. In the yrast SD band of this isotope we have identified a 
transition of 124 keV energy which is the lowest $\gamma$-ray energy observed in any SD nucleus with the exception of the
fission isomers. This SD band of $^{196}$Bi has been interpreted as built on the configuration
    ($\pi$[651]1/2$\otimes \nu$[752]5/2)$K^\pi=2^-$, and the 124 keV $\gamma$-ray is proposed to be 
    the 9$^-\rightarrow 7^-$ transition. 

\section*{Acknowledgments}

We would like to thank all those involved in the setting up and
commissioning of EUROBALL~IV. We are also very indebted to the operators of 
the VIVITRON tandem who have provided us a very stable $^{19}$F beam during the six days of the experiment. 
We would like to thank J. Meyer for enlightening and helpful discussions. The work of the Bonn Group was 
supported by BMBF contract no 06 BN 907.

\end{document}